\documentclass[aps,showpacs,preprintnumbers]{revtex4}%
\usepackage{graphicx}
\usepackage{array}
\usepackage{amsmath,amssymb}
\usepackage[latin1]{inputenc}
\usepackage{amsmath}
\usepackage{amsfonts}
\usepackage{amssymb}%
\providecommand{\U}[1]{\protect\rule{.1in}{.1in}}
\begin{document}
\title{INTERACTION OF FLEXURAL PHONONS WITH ELECTRONS IN GRAPHENE: A GENERALIZED
DIRAC EQUATION IN CORRUGATED SURFACES}
\author{Richard Kerner $^{1,*}$}
\author{Gerardo G Naumis$^{2}$}

\begin{abstract}
A generalized Dirac equation is derived in order to describe charge carriers
moving in corrugated graphene, which is the case for temperatures above $10%
{{}^\circ}%
K$ due to the presence of flexural phonons. Such interaction is taken into
account by considering an induced metric, in the same spirit as the general
relativity approach for the description of fermionic particle moving in a
curved space-time. The resulting equation allows to include in a natural way
the presence of other phonon branches as well as an external electromagnetic
field. It also predicts non-linear effects which are not present in the usual
vector potential approximation used in most of publications on the subject, as
well as the possibility of controlling electronic conductivity using pure
sinusoidal strain fields. The non-linear terms are important at high
temperatures, and can also lead to interesting effects, like e.g. resonances
between flexural phonons and external electromagnetic fields.

\end{abstract}
\maketitle
\affiliation{1. LPTMC, Universit\'e Pierre et Marie Curie - CNRS URMR 7600, Tour 23,
5-\`eme \'etage, Boite 121, 4 Place Jussieu, 75005 Paris, France.}
\affiliation{}
\affiliation{2. Depto. de Física-Química, Instituto de Física, Universidad Nacional
Aut\'onoma de M\'exico (UNAM). Apdo. Postal 20-364, 01000, M\'exico D.F., M\'exico.}

\section{Introduction}

Graphene is a new material that has been attracting a lot of attention since
its experimental discovery \cite{novoselov}. This carbon allotrope has unique
transport properties \cite{geimgra}\cite{peres}, like a high electronic
mobility \cite{novoselov2} and thermal conductivity \cite{balandin}, which are
\ believed to be important for future applications in nano-devices
\cite{avouris}\cite{CrestiReview}. However, there are certain discrepancies in
the values of the electronic mobilities depending on whether the samples are
suspended or in a substrate \cite{bolotin2008}\cite{castro&gorbachev2010}. At
low temperatures, impurity scattering can be responsible for this effect,
which eventually leads to a metal-insulator transition since the mobility edge
appears near the Fermi energy \cite{naumis}, as has been confirmed in graphene
doped with H \cite{bostwick}. However, above $T>10%
{{}^\circ}%
K$ such discrepancies are believed to be a consequence of the crucial role of
flexural phonon modes in the electron scattering, as has been shown very
recently by applying tension to graphene sheets \cite{castro&gorbachev2010}.
In fact, flexural modes are collective atomic displacements which are
perpendicular to the graphene's plane \cite{CastroNetoReview}, making it
behave like a corrugated surface. From a microscopic point of view, the
scattering results from changes in the distances between atoms, leading to
fluctuations of $\pi-$orbitals electron wavefuctions overlaps \cite{Ando2006}.

However, since charge carriers in flat graphene are described by massless
Dirac fermions \cite{novoselov2,katsnelson}, it is natural to ask if it is
possible to modify the Dirac equation taking into account the flexural mode
interaction. Two paths can be followed in order to answer this question. One
is to start from the usual tight-binding approach and use Taylor expansion of
the overlap integral on the displacement field \cite{CastroNetoReview}%
\cite{Ando2006}\cite{CastroGuineaElectronPhonon}. Here we present an
alternative point of view, in which the interaction is included by making the
observation that a graphene membrane can be considered as a curved space. The
effective equation must be covariant due to simple and general physical
arguments. The desired equation can be considered as akin to the Dirac
equation in curved space-time. In fact, a similar approach has been taken in
references \cite{VozmedianoEPJ}\cite{VozmedianoJPC}\cite{VozmedianoGuinea2010}
to study smooth ripples in graphene, using a metric that was considered
asymptotically flat and with cylindrical symmetry. Our approach does not
contain such requirements since the metric is general and thus can include a
spatially non-decaying phonon field. This approach leads to interesting
predictions, like the possibility of building gates for graphene electronic
devices by superposing sinusoidal strain fields in different directions. Also,
since the planar vibrational modes can be described by vector potential
\cite{CastroNetoReview}\cite{Ando2006}\cite{CastroGuineaElectronPhonon}, our
approach allows to include all phonon branches and the electromagnetic
potential in a single equation\textbf{.} Note that in fact, the problem of the
two dimensional Dirac equation including a vector potential has been solved
recently \cite{lopez}\cite{lopez2}. As we shall show in the conclusion, the
present approach has certain advantages over the Taylor expansion of the
tight-binding parameters.

To finish this introduction, let us briefly sketch the ideas behind the
present approach. For graphene at low temperatures, its surface can be
considered as flat. The corresponding unperturbed Hamiltonian operator used to
describe charges evolving on the graphene sheet can be written as an effective
Dirac equation \cite{novoselov2,katsnelson},
\begin{equation}
\hat{H}=-i\hslash v_{F}\left[  \gamma^{x}\,\nabla_{x}+\gamma^{y}\,\nabla
_{y}\right]  , \label{Hhat}%
\end{equation}
where $v_{F}$ is the \textit{Fermi velocity}, with
\begin{equation}
\gamma^{x}=\sigma_{x}=%
\begin{pmatrix}
0 & 1\cr1 & 0
\end{pmatrix}
,\;\;\;\gamma^{y}=\sigma_{y}=%
\begin{pmatrix}
0 & -i\cr i & 0
\end{pmatrix}
. \label{gammasigma}%
\end{equation}
Note that at this stage there is no difference between covariant and
contravariant indices, because we have
\begin{equation}
\gamma^{i}\gamma^{j}+\gamma^{j}\gamma^{i}=\sigma_{i}\sigma_{j}+\sigma
_{j}\sigma_{i}=2\,g_{ij}\,\mathbf{1},\;\;\;(i,j=x,y), \label{anticomm}%
\end{equation}
with $g_{ij}=\delta_{ij}$, so that obviously the contravariant metric raising
the indices is also $g^{ij}=\delta^{ij}$. But this is no more true when the
underlying two-dimensional space is not flat, but corrugated, with a non
trivially deformed metric, as the case is for graphene at $T>>10%
{{}^\circ}%
K$. This is why a more careful treatment of this two-dimensional version of
the Dirac equation should be considered. Luckily enough, the problem of
covariant formulation of Dirac's equation in a curved space has been quite
deeply investigated since a long time \cite{Lichnerowicz, Birrell}, so that we
can follow the same steps in this particular case: introduce the non-Euclidean
two-dimensional metric on the graphene sheet, then adapt the Clifford algebra
and find the Christoffel connection, and finally assemble all these in the
covariant version of Dirac's equation.

The layout of the present article is the following: Section II describes the
induced metric due to a corrugated surface, and in Section III we obtain the
resulting covariant Dirac equation. Section IV deals with the application of
the generalized Dirac equation to the description of electrons evolving on the
corrugated graphene sheet, and finally, conclusions are given in Section V.

\section{The induced metric}

At any finite temperature, phonons produce a displacement field
($\overrightarrow{u}$). We start in this section by finding the resulting
metrics and its Christoffel connection on corrugated graphene. Let the surface
materialized by the graphene sheet be described by,
\begin{equation}
z=f(x,y,t), \label{thesheet}%
\end{equation}
where $x,y$ are the coordinates in the graphene sheet, and $z$ is the out of
plane displacement. Notice that here we only consider the flexural phonon
branch, since the planar phonons can be described by inclusion in a vectorial
potential \cite{CastroNetoReview}, whose solution is basically known
\cite{lopez}\cite{lopez2}.

As now the differential $dz$ becomes a linear combination of $dx$ and $dy$, we have%

\begin{equation}
dz = \frac{\partial f}{\partial x} dx + \frac{\partial f}{\partial y} dy,
\label{diffz}%
\end{equation}

and the induced metric on the sheet is given by the following formula:
\begin{equation}
ds^{2}=g_{ij}\,dx^{i} dx^{j},\;\;\;(i,j = x,y,z). \label{metricabstract}%
\end{equation}
where $g_{ij}$ is the metric tensor. Or, in a more explicit form,
\begin{equation}
ds^{2}=dx^{2}+dy^{2}+dz^{2}=dx^{2}+dy^{2}+(\frac{\partial f}{\partial
x}dx+\frac{\partial f}{\partial y}dy)^{2}. \label{diffz2}%
\end{equation}
and it is obviously a metric in a $2$-dimensional (but curved) space
parametrized two space variables $(x,y)$ After opening the last expression we
get the explicit form of the induced metric, which is {\Large
\begin{equation}
{\tilde{g}}_{jk}=%
\begin{pmatrix}
1+(\frac{\partial f}{\partial x})^{2}\; & \frac{\partial f}{\partial x}%
\frac{\partial f}{\partial y}\cr\frac{\partial f}{\partial x}\frac{\partial
f}{\partial y} & 1+(\frac{\partial f}{\partial y})^{2}%
\end{pmatrix}
\label{metricmatrix}%
\end{equation}
} This can be written as
\begin{equation}
{\tilde{g}}_{jk}=g_{jk}+h_{jk},\;\;\mathrm{with}\;\;j,k=x,y.
\label{gdeformation}%
\end{equation}
where $g_{ij}=\mathrm{diag}(1,1)$ is the flat metric in $2$-dimensional space,
and $h_{ij}$ is the perturbation (supposed small as compared with $1$)
provoked by the corrugation of the sheet. The matrix (\ref{metricmatrix}) is
symmetric and real, therefore it can be diagonalized by an appropriate linear transformation.

Also the inverse (contravariant) metric can be easily found. The determinant
of the matrix corresponding to the covariant metric tensor (\ref{metricmatrix}%
) is easily found to be
\begin{equation}
\mathrm{det}\,%
\begin{pmatrix}
1+(\frac{\partial f}{\partial x})^{2}\; & \frac{\partial f}{\partial x}%
\frac{\partial f}{\partial y}\cr\frac{\partial f}{\partial x}\frac{\partial
f}{\partial y} & 1+(\frac{\partial f}{\partial y})^{2}%
\end{pmatrix}
=(1+(\partial_{x}f)^{2}+(\partial_{y}f)^{2}); \label{detgcov}%
\end{equation}
and the inverse matrix, corresponding to the contravariant metric $g^{jk}$ is
\begin{equation}
g^{jk}=\frac{1}{Q}\;%
\begin{pmatrix}
1+(\frac{\partial f}{\partial y})^{2}\; & -\frac{\partial f}{\partial x}%
\frac{\partial f}{\partial y}\cr-\frac{\partial f}{\partial x}\frac{\partial
f}{\partial y} & 1+(\frac{\partial f}{\partial x})^{2}%
\end{pmatrix}
\label{detgcov2}%
\end{equation}
where we used the abbreviate notation for the determinant, $Q=1+(\partial
_{x}f)^{2}+(\partial_{y}f)^{2}$.

The deformation $h_{jk}$ is entirely composed of quadratic terms containing
products of \textit{spatial} partial derivatives of the deformation function
$f(x,y,t)$,
\begin{equation}
h_{jk} \simeq\partial_{j} f \, \partial_{k} f,
\end{equation}
and it is easy to check that the corresponding Christoffel symbols reduce to
\begin{equation}
{\tilde{\Gamma}}^{i}_{jk} = \frac{1}{2} \, {\tilde{g}}^{im} \biggl( \partial
_{j} {\tilde{g}}_{mk} + \partial_{k} {\tilde{g}}_{jm} - \partial_{m}
{\tilde{g}}_{jk} \biggr) = {\tilde{g}}^{im} \partial_{m} f \, \partial
^{2}_{jk} f. \label{Christoffels1}%
\end{equation}
Both quantities disappear when $f = 0$, and the metric becomes flat again.

The Dirac equation in two dimensions should be now generalized in order to
incorporate the fact that the metric on the surface of constraint is no more
flat, but curved. In the following section, we obtain the covariant
generalization of the Dirac equation.

\section{The covariant Dirac equation}

The equation we want to produce now can be written as the new deformed
quantum-mechanical Hamiltonian acting on a two-component spinor $\Psi$ as follows:%

\begin{equation}
{\hat{\tilde{H}}}\,\Psi\sim\left[  {\tilde{\gamma}}^{x}\,{\tilde{\nabla}}%
_{x}+{\tilde{\gamma}}^{y}\,{\tilde{\nabla}}_{y}\right]  \,\Psi.
\label{Hhatexpl}%
\end{equation}
Here not only the contravariant metric is deformed, but also the $\gamma
$-matrices should be modified in order to satisfy new anti-commutation
relations with the induced metric instead of the flat one as before; finally,
${\tilde{\nabla}}_{j}$ contains not only the electromagnetic and on plane
phonon interaction visualized by the vector potential included in the usual
gauge-invariant way, but also the Christoffel symbols of the metric
${\tilde{g}}_{ij}$:
\begin{equation}
\nabla_{j}\,\Psi=(\partial_{j}-e\,A_{j})\Psi+{\tilde{\Gamma}}_{jk}%
^{m}\,{\tilde{g}}^{ki}\,{\tilde{\Sigma}}_{mk}\,\Psi, \label{covdiffspin}%
\end{equation}
where the Christoffel symbols ${\tilde{\Gamma}}_{jk}^{m}$ are defined as
usual, by means of the modified metric:
\begin{equation}
{\tilde{\Gamma}}_{jk}^{i}=\frac{1}{2}\,{\tilde{g}}^{im}\,[\partial_{j}%
{\tilde{g}}_{mk}+\partial_{k}{\tilde{g}}_{jm}-\partial_{m}{\tilde{g}}_{jk}\,],
\label{Christoffel}%
\end{equation}
and ${\tilde{\Sigma}}_{mk}$ is the matrix-valued anti-symmetric tensor defined
by means of the modified gamma-matrices:
\begin{equation}
{\tilde{\Sigma}}_{mk}=\frac{1}{8}[{\tilde{\gamma}}_{m}{\tilde{\gamma}}%
_{k}-{\tilde{\gamma}}_{k}{\tilde{\gamma}}_{m}]. \label{Sigmadef}%
\end{equation}
This term is often called \textquotedblleft spinorial connection".
\cite{Lichnerowicz, Birrell}.

We are looking for two ``deformed" generators of the Clifford algebra
${\tilde{\gamma}}_{x}\;\;\mathrm{and}\;\;{\tilde{\gamma}}_{y}$ that would
satisfy
\begin{equation}
{\tilde{\gamma}}_{x}{\tilde{\gamma}}_{x}=-\biggl(\frac{\partial f}{\partial
x}\biggr)^{2}\,\mathbf{1},\,\;\;\;{\tilde{\gamma}}_{x}{\tilde{\gamma}}%
_{y}+{\tilde{\gamma}}_{y}{\tilde{\gamma}}_{x}=-2\,\frac{\partial f}{\partial
x}\frac{\partial f}{\partial y}\,\mathbf{1},\,\;\;\;{\tilde{\gamma}}%
_{y}{\tilde{\gamma}}_{y}=-\biggl(\frac{\partial f}{\partial y}\biggr)^{2}%
\,\mathbf{1}. \label{newcommut}%
\end{equation}
In order to do this, we must introduce the third Pauli matrix, because the
deformation of the sheet pushes it out of the strict two-dimensional plane
$(x,y)$.

The undeformed Clifford algebra satisfying anti-commutation relations in a
three-dimensional \textit{flat} space is defined as follows:
\begin{equation}
\gamma_{j} \gamma_{k} + \gamma_{k} \gamma_{j} = 2 g_{jk} \, \mathbf{1} \; \;
\mathrm{with} \, \; \; g_{jk} = \mathrm{diag} (1, 1, 1), \; \; j, k = 1,2,3,
\label{Cliffgdiag}%
\end{equation}
is easily found to be generated by three Pauli matrices as follows:
\begin{equation}
\gamma_{1} = \sigma_{x}, \; \; \; \gamma_{2} = \sigma_{y}, \; \; \gamma_{3} =
\sigma_{z}, \label{gammasigma2}%
\end{equation}
with
\begin{equation}
\sigma_{x} =
\begin{pmatrix}
0 & 1 \cr 1 & 0
\end{pmatrix}
, \; \; \sigma_{y} =
\begin{pmatrix}
0 & -i \cr i & 0
\end{pmatrix}
, \sigma_{z} =
\begin{pmatrix}
1 & 0 \cr 0 & -1
\end{pmatrix}
, \; \; \label{Paulisigmas}%
\end{equation}
we have indeed
\[
(\gamma_{x})^{2} = \mathbf{1}, \; \; (\gamma_{y})^{2} = \mathbf{1}, \; \; \;
(\gamma_{z})^{2} = \mathbf{1}%
\]
and the three matrices anticommuting with each other.

The ansatz for the two deformed space-like $\gamma$-matrices is simple: if we
set
\begin{equation}
{\tilde{\gamma}}_{x}=\sigma_{x}+a\sigma_{z},\,\;\;\;{\tilde{\gamma}}%
_{y}=\sigma_{y}+b\sigma_{z}, \label{gammaansatz}%
\end{equation}
then the coefficients $a$ and $b$ should be, as it was easy to check,
\begin{equation}
a=\frac{\partial f}{\partial x},\;\;\;\;b=\frac{\partial f}{\partial y},
\label{abcoeffs}%
\end{equation}
so that
\begin{equation}
{\tilde{\gamma}}_{x}=\sigma_{x}+\frac{\partial f}{\partial x}\,\sigma
_{z},\,\;\;\;{\tilde{\gamma}}_{y}=\sigma_{y}+\frac{\partial f}{\partial
y}\,\sigma_{z}, \label{spacegammas}%
\end{equation}
Now we have to produce their \textit{contravariant} counterparts that appear
in the Dirac equation \cite{Birrell}. We have:
\begin{equation}
{\tilde{\gamma}}^{x}={\tilde{g}}^{xx}\,{\tilde{\gamma}}_{x}+{\tilde{g}}%
^{xy}\,{\tilde{\gamma}}_{y},\;\;\;{\tilde{\gamma}}^{y}={\tilde{g}}%
^{yx}\,{\tilde{\gamma}}_{x}+{\tilde{g}}^{yy}\,{\tilde{\gamma}}_{y},
\label{contragammas}%
\end{equation}
which gives explicitly
\[
{\tilde{\gamma}}^{x}=\frac{1}{Q}\,\Biggl[\biggl(1+(\partial_{y}f)^{2}%
\biggr)\sigma_{x}+(\partial_{x}f)\,\sigma_{z}-(\partial_{x}f)(\partial
_{y}f)\,\sigma_{y},\Biggr],
\]%
\begin{equation}
{\tilde{\gamma}}^{y}=\frac{1}{Q}\,\Biggl[\biggl(1+(\partial_{x}f)^{2}%
\biggr)\sigma_{y}+(\partial_{y}f)\,\sigma_{z}-(\partial_{x}f)(\partial
_{y}f)\,\sigma_{x}\Biggr].
\end{equation}
Recalling that $Q=1+(\partial_{x}f)^{2}+(\partial_{y}f)^{2}$, we can add and
substract in the numerators of the above formula respectively the following
terms: $(\partial_{x}f)^{2}\,\sigma_{x}$ in the first one, and $(\partial
_{y}f)^{2}\,\sigma_{y}$ in the second one; this will enable us to separate the
undeformed matrices $\sigma_{x}$ and $\sigma_{y}$ and the genuine deformation
terms containing spatial derivatives of $f$. This gives the following result:
\begin{equation}
{\tilde{\hat{H}}}\sim\sigma^{x}\,{\tilde{\nabla}}_{x}+\sigma^{y}%
\,{\tilde{\nabla}}_{y}+\frac{\sigma_{z}}{Q}\,({\vec{\mathrm{grad}}}f\cdot
{\vec{\nabla}})-\frac{1}{Q}({\vec{\mathrm{grad}}}f\cdot{\vec{\sigma}}%
)({\vec{\mathrm{grad}}}f\cdot{\vec{\nabla}}) \label{Htildeexpl}%
\end{equation}
where the vectors and their scalar products are two-dimensional, i.e. we
mean:
\[
{\vec{\mathrm{grad}}}f=[\partial_{x}f,\;\partial_{y}\,f],\;\;{\vec{\sigma}%
}=[\sigma_{x},\,\sigma_{y}\,],\;\;\;{\vec{\nabla}}=[\nabla_{x},\,\nabla_{y}],
\]
so that
\[
(\vec{\mathrm{grad}}f)\cdot(\vec{\sigma})=\partial_{x}f\,\sigma_{x}%
+\partial_{y}f\,\sigma_{y},\;\;\;\mathrm{etc.}%
\]
We see that already in the numerators we have not only linear, but also
quadratic terms, notwithstanding the presence of quadratic terms in the
denominator (contained in the normalizing factor $Q$). If we decided to keep
\textit{linear terms only}, then the modified Hamiltonian would contain only
one extra term proportional to the matrix $\sigma_{z}$:
\begin{equation}
{\tilde{\hat{H}}}_{lin}\sim{\hat{H}}_{0}+\delta{\hat{H}=\sigma^{x}\,{\nabla
}_{x}+\sigma^{y}\,\nabla}_{y}+\sigma_{z}\,({\vec{\mathrm{grad}}}f\cdot
{\vec{\nabla}}) \label{Hlinear}%
\end{equation}
Note that also the differential operator $\nabla$ is taken in its primary
form, because the connection coefficients contain only quadratic expressions
in derivatives of $f$. Observe that this equation is akin to the one obtained
using a Taylor expansion of the overlap integral in a tight-binding approach.

However, if we choose to keep all terms up to quadratic ones, then we must
take into account also the Christoffel coefficients in ${\tilde{\nabla}}$. It
is easy to check the following explicit form of our Christoffel symbols;
keeping only the second order expressions means that we can use the simplified
formula in which ${\tilde{g}}^{ij}$ is replaced by $g^{ij}=\delta^{ij}$. We
have then:
\[
\Gamma_{xx}^{x}=\partial_{x}f\,\partial_{xx}^{2}f,\;\;\;\Gamma_{xy}^{x}%
=\Gamma_{yx}^{x}=\partial_{x}f\,\partial_{xy}^{2}f,\;\;\;\Gamma_{xx}%
^{x}=\partial_{x}f\,\partial_{yy}^{2}f,
\]%
\begin{equation}
\Gamma_{xx}^{y}=\partial_{y}f\,\partial_{xx}^{2}f,\;\;\;\Gamma_{xy}^{y}%
=\Gamma_{yx}^{y}=\partial_{y}f\,\partial_{xy}^{2}f,\;\;\;\Gamma_{yy}%
^{y}=\partial_{y}f\,\partial_{yy}^{2}f,\label{Christoffelex}%
\end{equation}
In covariant derivatives, these coefficients are contracted with $g^{jk}$ and
the \textit{anti-symmetric} matrices $\Sigma_{km}$,
\[
\Gamma_{jk}^{i}\,g^{km}\,\Sigma_{im},
\]
which, taking into account that only diagonal terms in $g^{ik}$ do not vanish
and are equal to one, leads to the following result when explicited: for $j=x$
we have
\[
\Gamma_{xx}^{x}g^{xx}\,\Sigma_{xx}+\Gamma_{xy}^{x}g^{yy}\,\Sigma_{xy}%
+\Gamma_{xx}^{y}g^{xx}\,\Sigma_{yx}+\Gamma_{xy}^{y}g^{yy}\,\Sigma_{yy},
\]
and for $j=y$ we get a similar expression:
\[
\Gamma_{yx}^{x}g^{xx}\,\Sigma_{xx}+\Gamma_{yy}^{x}g^{yy}\,\Sigma_{xy}%
+\Gamma_{yx}^{y}g^{xx}\,\Sigma_{yx}+\Gamma_{yy}^{y}g^{yy}\,\Sigma_{yy},
\]
The non-vanishing metric tensor components are equal to one, whereas the
$\Sigma$-matrices are anti-symmetric in their two lower indices, so what is
left is only
\[
\Gamma_{xy}^{x}\,\Sigma_{xy}+\Gamma_{xx}^{y}\,\Sigma_{yx}=\biggl(\partial
_{x}f\,\partial_{xy}^{2}f-\partial_{y}f\,\partial_{xx}^{2}f\biggr)\Sigma
_{xy}=-(\partial_{y}f\,\partial_{xx}^{2}f)\Sigma_{xy},
\]%
\begin{equation}
\Gamma_{yy}^{x}\,\Sigma_{xy}+\Gamma_{yx}^{y}\,\Sigma_{yx}=\biggl(\partial
_{y}f\,\partial_{xx}^{2}f-\partial_{x}f\,\partial_{yy}^{2}f\biggr)\Sigma
_{xy}=-(\partial_{x}f\,\partial_{yy}^{2}f)\Sigma_{yx}\label{withsigmas}%
\end{equation}
These are the only terms remaining to be included in the covariant derivatives
as follows:
\[
{\tilde{\nabla}}_{x}=(\partial_{x}-eA_{x})-(\partial_{y}f\,\partial_{xx}%
^{2}f)\Sigma_{xy},\;\;\;{\tilde{\nabla}}_{y}=(\partial_{y}-eA_{y}%
)-(\partial_{x}f\,\partial_{yy}^{2}f)\Sigma_{xy}.
\]
In the Hamiltonian, they appear multiplied from the left by the corresponding
${\tilde{\gamma}}$-matrices, but here, evaluating the terms coming from the
Christoffel connection coefficients, already quadratic in deformation $f$, we
may keep only their undeformed version, which our case are just the two Pauli
matrices $\sigma^{x}$ and $\sigma^{y}$, so that the part of the deformed
Hamiltonian keeping track of the Christoffel connection is
\[
\sigma^{x}\,{\tilde{\nabla}}_{x}+\sigma^{y}\,{\tilde{\nabla}}_{y}=\sigma
^{x}\,(\partial_{x}-eA_{x})+\sigma^{x}\,\biggl[(\partial_{x}f)(\partial
_{xy}^{2}f)-(\partial_{y}f)\,(\partial_{xx}^{2}f)\biggr]\,\Sigma_{xy},
\]%
\begin{equation}
+\sigma^{y}\,(\partial_{y}-eA_{y})+\sigma^{y}\,\biggl[(\partial_{y}%
f)(\partial_{yx}^{2}f)-(\partial_{x}f)\,(\partial_{yy}^{2}f))\biggr]\,\Sigma
_{yx}.\label{Hamchristoffel}%
\end{equation}
Taking into account the fact that
\[
(\sigma_{x})^{2}=\mathbf{1},\;\;\;(\sigma_{y})^{2}=\mathbf{1},\;\;\;\sigma
_{x}\sigma_{y}=-\sigma_{y}\sigma_{x},\;\;\mathrm{and}\;\;\;\sigma^{x}%
\Sigma_{xy}=\frac{1}{4}\,\sigma_{y},
\]
and adding and substracting terms like $(\partial_{y}f)(\partial_{yy}^{2}f)$
to the first expression and $(\partial_{x}f)(\partial_{xx}^{2}f)$ to the
second, we get the following invariant form of the extra terms induced by the
Christoffel connection:
\[
\frac{1}{4}\,\sigma^{y}\biggl[(\partial_{x}f)(\partial_{xy}^{2}f)+(\partial
_{y}f)(\partial_{yy}^{2}f)-(\partial_{y}f)(\partial_{yy}^{2}f)-(\partial
_{y}f)\,(\partial_{xx}^{2}f)\biggr]
\]%
\begin{equation}
+\frac{1}{4}\,\sigma^{x}\,\biggl[(\partial_{y}f)\,(\partial_{xy}%
^{2}f)+(\partial_{x}f)(\partial_{xx}^{2}f)-(\partial_{x}f)(\partial_{xx}%
^{2}f)-(\partial_{x}f)(\,\partial_{yy}^{2}f)\biggr]\label{withextras}%
\end{equation}
This in turn can be written in a more compact (and elegant !) way as follows:
\begin{equation}
\frac{1}{8}\,{\vec{\sigma}}\cdot{\vec{\mathrm{grad}}}\left[  (\partial
_{x}f)^{2}+(\partial_{y}f)^{2}\right]  -\frac{1}{4}\,\left[  {\vec{\sigma}%
}\cdot{\vec{\mathrm{grad}}}f\right]  \,\Delta f\label{elegant}%
\end{equation}
with
\[
\Delta f=\partial_{xx}^{2}f+\partial_{yy}^{2}f;
\]
It is worthwhile to note that the expression (\ref{elegant}) vanishes when $f$
is a pure monochromatic wave, $f=A\cos(\omega t-K_{x}x-K_{y}y),$ but is
different from zero as soon as there is a superposition of such expressions,
e.g. for a standing wave. It is wortwhile to mention that this simple fact
provides a nice prediction of the present approach, since it means that one
can control the electronic conductivity by mixing sinusoidal strain fields.
For example, if two sinusoidal strains are applied in perpendicular directions
using piezoelectric nano-devices, this will result in a very different
electronic conductivity when compared with the case of only one sinusoidal
deformation, due to the presence of the expression (\ref{elegant}). The reason
of such effect can be understood by observing that a pure sinusoidal
deformation does not change distances inside the sheet, as can be easily
verified by applying strain to a sheet of paper. If the strain is applied in
only one direction, the sheet can be gently bended, while a second strain in a
perpendicular direction produces real corrugation. 

Now we are able to write down the full Hamiltonian for an electron on a sheet,
taking into account that sheet's proper motions described by the deformation
from horizontal plane given by $z=f(x,y,t)$, up to the second order (quadratic
terms in derivatives of $f$):%

\[
{\hat{\tilde{H}}} \sim{\hat{H}}_{0} + \frac{1}{Q} \, \sigma_{z} \,
({\vec{\mathrm{grad}}}f) \cdot(\vec{\nabla}) - \frac{1}{Q} \, ({\vec
{\mathrm{grad}}} f \cdot{\vec{\sigma}})( {\vec{\mathrm{grad}}} f \cdot
\vec{\nabla})
\]
\begin{equation}
+\frac{1}{8} \, {\vec{\sigma}} \cdot{\vec{\mathrm{grad}}} \left[
(\partial_{x} f)^{2} + (\partial_{y} f)^{2} \right]  - \frac{1}{4} \, \left[
{\vec{\sigma}} \cdot{\vec{\mathrm{grad}}} f \right]  \, \Delta f.
\end{equation}
The normalizing factor $1/Q$ in front of two first contributions can be set to
$1$, because it contains the squares of derivatives of $f$, and if developed,
will create terms of order $3$ and $4$ when multiplied by the terms behind.

The quantum-mechanical Hamiltonian is obtained from this expression by
multiplying it by $-i\hbar$ and $v_{F}$. Let us define the operator of
generalized momentum,
\[
{\hat{\vec{\pi}}}={\hat{\vec{p}}}-e{\vec{A}}=-i\hbar\;{\vec{\mathrm{grad}}%
}-e{\vec{A}}%
\]
where ${\vec{A}}$ is a vector potential that describes an electromagnetic
field \cite{lopez}\cite{lopez2} or in plane longitudinal and transversal
phonons \cite{CastroGuineaElectronPhonon}. Then we can write:
\[
{\tilde{\hat{H}}}=v_{F}%
\begin{pmatrix}
0 & {\hat{\pi}}_{x}-i{\hat{\pi}}_{y}\cr {\hat{\pi}}_{x}+i{\hat{\pi}}_{y} & 0
\end{pmatrix}
+v_{F}%
\begin{pmatrix}
(\partial_{x}f)\,{\hat{\pi}}_{x}+(\partial_{y}f)\,{\hat{\pi}}_{y} & 0\cr0 &
-(\partial_{x}f)\,{\hat{\pi}}_{x}-(\partial_{y}f)\,{\hat{\pi}}_{y}%
\end{pmatrix}
\]%
\begin{equation}
-i\hbar v_{F}\left[  ({\vec{\mathrm{grad}}}f\cdot{\vec{\sigma}})({\vec
{\mathrm{grad}}}f\cdot\vec{\nabla})+\frac{1}{8}\,{\vec{\sigma}}\cdot
{\vec{\mathrm{grad}}}\left[  (\partial_{x}f)^{2}+(\partial_{y}f)^{2}\right]
-\frac{1}{4}\,\left[  {\vec{\sigma}}\cdot{\vec{\mathrm{grad}}}f\right]
\,\Delta f\right]  . \label{PseudoDirac}%
\end{equation}
This last equation is the generalized Dirac equation. For graphene, it is
interesting to observe that in the Hamiltonian derived here, the terms
${\vec{\sigma}}\cdot{\vec{\mathrm{grad}}}\left[  (\partial_{x}f)^{2}%
+(\partial_{y}f)^{2}\right]  /8-$ $\left[  {\vec{\sigma}}\cdot{\vec
{\mathrm{grad}}}f\right]  \,\Delta f/4$ are akin to the ones obtained from a
vectorial potential approach \cite{CastroNetoReview}. However, Eq.
(\ref{PseudoDirac}) contains more terms, including a dominant linear
correction in $f$.

\section{The generalized Dirac equation on corrugated graphene}

To show how Eq. (\ref{PseudoDirac}) is used in the graphene case, we consider
that the displacement field can be written using a simple set of basis
functions provided by standing waves \cite{kittel},%
\begin{equation}
\overrightarrow{u}=\frac{2}{\sqrt{N}}%
{\displaystyle\sum\limits_{\mu,\mathbf{q}>0}}
\overrightarrow{e}_{\mu}(\overrightarrow{q})\left[  Q_{\mu,\overrightarrow{q}%
}^{(c)}\cos\left(  \overrightarrow{q}\mathbf{\cdot}\overrightarrow
{R}\mathbf{-}\omega_{\mu}(\overrightarrow{q})t\right)  +Q_{\mu,\overrightarrow
{q}}^{(s)}\sin\left(  \overrightarrow{q}\mathbf{\cdot}\overrightarrow
{R}\mathbf{-}\omega_{\mu}(\overrightarrow{q})t\right)  \right]
\end{equation}
where $\mathbf{e}_{\mu}(\mathbf{q})$ is the polarization vector for a
wave-vector $\overrightarrow{q}$, $\mu$ is the phononic branch, $\omega_{\mu
}(\overrightarrow{q})$ the dispersion relationship and $\overrightarrow
{R}=(x,y)$ is the position. Notice that the notation $\mathbf{q}>0$ indicates
that $q_{x}>0$ and $q_{y}>0$. $Q_{\mu,\overrightarrow{q}}^{(c)}$ and
$Q_{\mu,\overrightarrow{q}}^{(s)}$ are operators given in terms of the
creation (annihilation) phonon operators \cite{kittel} $a_{\mu,\overrightarrow
{q}}$ ($a_{\mu,\overrightarrow{q}}^{\dag}$),
\begin{equation}
Q_{\mu,\overrightarrow{q}}^{(\alpha)}=\frac{1}{\sqrt{2M\omega_{\mu
}(\overrightarrow{q})}}\left(  a_{\mu,\overrightarrow{q}}^{(\alpha)}%
+a_{\mu,\overrightarrow{q}}^{(\alpha)\dag}\right)
\end{equation}
where $\alpha$ runs over $c$ and $s$ and $M$ is the carbon atom's mass.

Thus, for flexural phonons the function $f(x,y,t)$ is given by,%
\begin{equation}
f(x,y,t)=\frac{2}{\sqrt{N}}%
{\displaystyle\sum\limits_{\mathbf{q}>0}}
\left[  Q_{F,\overrightarrow{q}}^{(c)}\cos\phi_{\mathbf{q}}%
+Q_{F,\overrightarrow{q}}^{(s)}\sin\phi_{\mathbf{q}}\right]
\end{equation}
where $\mu=F$ means that we are dealing with flexural phonons, and the phase
$\phi_{\mathbf{q}}$ is defined as $\phi_{\mathbf{q}}=\overrightarrow
{q}\mathbf{\cdot}\overrightarrow{R}\mathbf{-}\omega_{\mu}(\overrightarrow
{q})t$. For the flexural branch, $\omega_{F}(\overrightarrow{q})=\alpha
_{F}\left\Vert \overrightarrow{q}\right\Vert ^{2}$ where $\alpha_{F}%
\approx4.6\times10^{-7}m^{2}/s$. Notice that $Q_{F,\overrightarrow{q}}^{(c)}$
and $Q_{F,\overrightarrow{q}}^{(s)}$ have length units, so $f(x,y,t)$ has the
same units, while $\partial_{x}f$ and $\partial_{y}f$ are dimensionless.

The generalized Dirac equation only needs as input the partial derivatives of
$f(x,y,t)$,%
\begin{align}
\partial_{x}f  &  =\frac{2}{\sqrt{N}}%
{\displaystyle\sum\limits_{\mathbf{q}>0}}
q_{x}\left[  Q_{F,\overrightarrow{q}}^{(c)}\cos\phi_{\mathbf{q}}%
+Q_{F,\overrightarrow{q}}^{(s)}\sin\phi_{\mathbf{q}}\right]  \text{,}\\
\text{ }\partial_{y}f  &  =\frac{2}{\sqrt{N}}%
{\displaystyle\sum\limits_{\mathbf{q}>0}}
q_{y}\left[  Q_{F,\overrightarrow{q}}^{(c)}\cos\phi_{\mathbf{q}}%
+Q_{F,\overrightarrow{q}}^{(s)}\sin\phi_{\mathbf{q}}\right]
\end{align}
For example, if we keep only the linear correction in Eq. (\ref{PseudoDirac}),
the Hamiltonian reads as follows,%

\[
{\tilde{\hat{H}}}=v_{F}%
\begin{pmatrix}
0 & {\hat{\pi}}_{x}-i{\hat{\pi}}_{y}\cr {\hat{\pi}}_{x}+i{\hat{\pi}}_{y} & 0
\end{pmatrix}
+\frac{2v_{F}}{\sqrt{N}}%
{\displaystyle\sum\limits_{\mathbf{q}>0}}
\left[  Q_{F,\overrightarrow{q}}^{(c)}\cos\phi_{\mathbf{q}}%
+Q_{F,\overrightarrow{q}}^{(s)}\sin\phi_{\mathbf{q}}\right]  \times
\]%
\begin{equation}%
\begin{pmatrix}
q_{x}\,{\hat{\pi}}_{x}+q_{y}\,{\hat{\pi}}_{y} & 0\cr0 & -q_{x}\,{\hat{\pi}%
}_{x}-q_{y}\,{\hat{\pi}}_{y}%
\end{pmatrix}
. \label{LinearDirac}%
\end{equation}

Using the fact that the Fermi velocity is much higher than the flexural modes'
speeds, we can drop the time dependence and apply perturbation theory to solve
the equation, as will be shown in the forthcoming article.

The correction terms that results from $\partial_{x}f$ and $\partial_{y}f$ are
of the order $z_{\max}/\lambda$, where $z_{\max}/\lambda$ is the maximal
vertical displacement of flexural mode, and $\lambda$ is the wavelength of a
typical ripple, usually around several atomic spacings $a_{0}$. Thus, the
energy of flexural modes correction behaves as $\sim v_{F}p_{F}(z_{\max}%
/a_{0})$ in the linear approximation, the subsequent non-linear terms being
proportional to the corresponding powers of the adimensional parameter
$z_{\max}/a_{0}$, as obtained in other approaches \cite{CastroNetoReview}.

\section{Conclusions}

We have deduced a generalized Dirac equation that describes charge carriers
moving in corrugated graphene. To produce such equation, an appropriate metric
was found and the principle of covariance has been applied. The resulting
equation contains a linear correction plus several non-linear terms. Some of
these non-linear terms correspond to the Taylor expansion of the overlap
integral approach\cite{CastroNetoReview}, while others, including the linear
correction, are new. Such terms could lead to interesting effects, like
resonances between various phonon modes, or between flexural phonons and the
oscillating external electromagnetic field. We expect these non-linear terms
to become important at higher temperatures, since the dependence of the
amplitude of $f(x,y,t)$ at higher temperatures contains the square root of
$T$, as can be easily shown by using the equipartition of energy principle.
Also, we proposed construction of simple electronic gate by observing that a
pure sinusoidal strain field has a very different effect on the electron
dynamics than a superposition of sinusoidals propagating in different directions.

\acknowledgments
One of us (RK) would like to thank M. Dubois-Violette for useful discussions
and remarks. The work was partly performed under the DGAPA-UNAM project IN-1003310-3.

\bibliographystyle{apsrev}
\bibliography{biblioImpEff}

\end{document}